\documentclass[12pt]{article}

\begin{document}
\setcounter{page}{1} \pagestyle{plain} \vspace{1cm}
\begin{center}
\Large{\bf Bound States and Many-Body Effects in H-Shaped Quantum
Wires}\\
\small \vspace{2cm} {\bf Kourosh Nozari}\\
\vspace{0.5cm}

{\it Department of Physics, Faculty of Basic Sciences,
 University of Mazandaran,\\
P. O. Box 47416-1467, Babolsar, IRAN\\
E-Mail: knozari@umz.ac.ir}\\
\vspace{0.5cm}
{\bf Mehrnoush Mirzaie}\\
\vspace{0.5cm}
 {\it Department of Physics, Iran
University of Science and
Technology, Tehran, IRAN\\
E-Mail:  mehrnoush@physics.iust.ac.ir}

\end{center}
\vspace{1.5cm}

\begin{abstract}
In this paper, bound states energies and corresponding wave
functions of H-shaped quantum wires are calculated numerically in
the presence of external magnetic and electric fields and within the
Landau gauge. With a suitable definition of external confinement
potential, we present a numerical algorithm to calculate the profile
of probability distribution of charge carriers. Our analysis shows
that in the presence of external electric and magnetic fields, bound
state properties of carriers are sensitive functions of an
asymmetric parameter $a=\frac{W_{x}}{W_{y}}$ which measures the
relative width of the well in two directions. We also study many
body effect of bandgap renormalization in this quasi one dimensional
system within dynamical random phase approximation.
\end{abstract}
\newpage

\section{Introduction}
A highly dense electron-hole plasma can be generated in a wide
variety of semiconductors by optical pumping. The band structure and
the optical properties of highly excited semiconductors generally
differ from those calculated for non-interacting electron-hole pairs
due to many-body exchange-correlation effects arising from the
electron-hole plasma[1,2]. In recent years, quasi-one dimensional
semiconductor quantum wires (QW) have been fabricated in a variety
of geometric shapes with atomic scale definition, and QW optical
properties have been studied for their potential device applications
such as semiconductor lasers[3,4]. Recently different geometries of
quantum wires, such as rectangular, V-shaped, L-shaped and T-shaped
quantum wires have been fabricated and some of their electronic
properties have been studied[4]. In addition, various experimental
techniques for fabrication and growth of these structures have been
developed[3-5]. Square quantum well wires have been studied by Hu
and Das Sarma[6]. They have calculated the value of the band gap
re-normalization( renormalization of fundamental gap structure due
to many body exchange-correlation effects) for this case within GW
approximation. Excitonic effects in quantum wires have been studied
by Goldoni {\it et al} [7]. They have studied the effects of Coulomb
interaction on the linear and nonlinear optical properties of both
V-shaped and T-shaped semiconductor quantum wires. Wang and Das
Sarma have proposed an elegant framework for numerical studies of
carrier induced many-body effects on the excitonic optical
properties of highly photoexcited one-dimensional quantum wire
systems[8,9]. Hwang and Das Sarma have investigated the dynamical
self-energy corrections of electron-hole plasma due to
electron-electron and electron-phonon interactions at the band edges
of a quasi-one dimensional photoexcited electron-hole plasma within
GW approximation[10,11]. Rinaldi and Cingolani have studied the
optical properties of quasi one dimensional quantum structures
specially the case of V-shaped quantum wires[12]. Bener and Haug
have considered plasma-density dependence of the optical spectra for
quasi-one-dimensional quantum well wires[13]. Tanatar has studied
the band gap re-normalization in quasi-one dimensional systems in a
simple plasmon-pole (quasi-static) approximation[14,15]. G\"{u}ven
{\it et al} have studied the band-gap renormalization in quantum
wire system within dynamical correlations and multi-subband
effects[16]. Luttinger liquid behavior of a semiconductor quantum
wire has been studied by Bellucci and Onorato[17]. They have also
studied the effects of magnetic field on low dimensional electron
systems focusing on Luttinger liquid behavior in a quantum wire[18].
In addition, they have studied also the ballistic electron transport
in a quantum wire under the action of an electric field[19] and
quenching of the spin Hall effect in ballistic nanojunctions[20]. As
an application, recently they have studied the transport through a
double barrier in large radius Carbon Nanotube with transverse
magnetic field[21]. Many particle aspects of a semiconductor quantum
wire within an improved random phase approximation has been studied
by Ashraf and Sharma[22]. They have considered structure factor,
pair distribution function, screened impurity potential and density
of screening charge and exchange and screened exchange energies
within an improved random phase approximation.

On the other hand, T-shaped and L-shaped quantum wires have been
studied by some authors recently. For example, Sedlmaier and his
coworkers have studied the band gap re-normalization of modulation
doped T-shaped quantum wires. They have presented a self-consistent
electronic structure calculation for this device[23]. Using a
density functional theory, Stopa has calculated the electronic
structure of a modulation doped and gated T-shaped quantum wire
[24]. Szymanska {\it et al} have studied the excitons in T-shaped
quantum wires[25]. They have calculated energies and  oscillator
strength for radiative recombination and two particle wave functions
for ground state exciton in a T-shaped quantum wire. Lin, Chen and
Chuu have found the dependence of the bound states of L-shaped and
T-shaped quantum wires to some asymmetric parameter in an
inhomogeneous magnetic field[26]. They have proposed a simple model
to explain the behavior of the magnetic dependence of the bound
state energy both in week and strong field regions. Recently Nozari
and Madadi have studied numerically the bound states properties and
the band gap renormalization of V-shaped and T-Shaped quantum wires
within dynamical random phase approximation[27,28]. Ballistic
transport through coupled T-shaped quantum wires has been studied by
Lin {\it et al} [29].

As another possible geometry of low dimensional systems, H-shaped
electronic systems have been considered and their many-body
electronic properties have been studied by some authors. Shin and
coworkers have studied quantum transport in an H-shaped quantum wire
and a ring structure[30]. They have studied numerically the
transport properties of an H-shaped quantum wire structure by using
the mode matching technique. They have reported the existence of
anomalous Hall resistance plateaus in this structure with relatively
low magnetic fields as precursors of integer Hall plateaus.
Henkiewicz {\it et al} have studied the manifestation of the spin
Hall effect in a two-dimensional electronic system with Rashba
spin-orbit coupling via dc-transport measurements in a realistic
H-shaped structure[31]. Designing of H-shaped micromechanical
structure has been studied by Arhaug and Soeraasen[32]. Our
investigation shows that there is no other concrete study of these
structure in existing literature. Obviously an analytical-numerical
study of this special structure is important to fill the existing
gap. Specially, our investigation of literature shows that bound
states of H-shaped quantum wires and many-body effects such as
bandgap renormalization of these low dimensional systems have not
been studied yet. So, in this paper we consider the geometry of
H-shaped quantum wires and by a suitable analytical definition of
quasi-one dimensional H-shaped confinement potential, we propose a
numerical scheme for calculating the bound states energies and wave
functions of charge carriers in the presence of external electric
and magnetic fields and within the Landau gauge. We obtain the
profile of charge carriers distribution(probability distribution) in
the presence of electric and magnetic fields. As an important
nonlinear optical effect, the many-body exchange-correlation induced
band gap narrowing in this type of quantum wire will be studied
within leading order dynamical random phase approximation.

The paper is organized as follows: section $2$ is devoted to
formulation of the problem and definition of confinement potential
analytically. In section $3$ we give a short but complete review of
bandgap renormalization in a general quasi-one dimensional
semiconductor. Some numerical details are given at the end of this
section. Section $4$ provides numerical results of our study and
their interpretation, while the numerical scheme of our calculations
based on finite difference algorithm is presented in the Appendix.
The results of each step are shown by figures. Finally, summary and
conclusions are given in section $5$.

\section{ The Setup}
The geometry of a typical H-shaped quantum wire is shown in figure
$1$. The typical values of $W_{x}$ and $W_{y}$ are of the order of
$5-50$ Nanometer. We study bound states and many-body effects in
this quasi one dimensional system in the presence of external
electric and magnetic fields. The electric field is assumed to be
directed along the $x$-axis and its typical value is of the order of
a few $V/cm$. The presence of magnetic field is in such a way that
the Landau gauge, $\vec{A}=(0,\,Bx,\,0)$ is satisfied. The
Hamiltonian of an electron in this configuration can be written as
follows
\begin{equation}
H=\frac{1}{2m}\Big[p_{x}^{2}+(p_{y}-eBx)^{2}\Big]-eEx
\end{equation}
where $e$ and $m$ are electron charge and mass respectively. The
Schr\"{o}dinger equation of this electron with wave function
$\Psi(x,y)$ can be written as follows
\begin{equation}
-\Bigg(\frac{\partial^{2}}{\partial
x^{2}}+\frac{\partial^{2}}{\partial
y^{2}}-\frac{2ieB}{\hbar}x\frac{\partial}{\partial
y}\Bigg)\Psi(x,y)+\Bigg(\frac{eB}{\hbar}\Bigg)^{2}x^{2}\Psi(x,y)-
\frac{2meE}{\hbar^{2}}x\Psi(x,y)=\frac{2m}{\hbar^{2}}\lambda\Psi(x,y)
\end{equation}
where $\lambda$ stands for eigenvalues of energy, $E$ represents the
value of electric field and $B$ is the value of the magnetic field.
To solve this eigenvalue problem we need the boundary conditions
which are given by the external confinement potential. The geometry
of H-shaped quantum wire as shown in figure $1$, suggests the
following definition of confinement potential
\begin{equation}
\label{math:1.1} V(x,y)=\left\{\begin{array}{ll} 0&{\rm
if}\left\{\begin{array}{ll}-\infty<x<+\infty,\quad-\frac{3}{2}W_{y}\leq
y\leq -W_{y}\\ -\infty<x<+\infty,\quad W_{y}\leq y\leq
\frac{3}{2}W_{y}\\
-\frac{W_{x}}{2}\leq x\leq\frac{W_{x}}{2},\quad
-W_{y}\leq y \leq W_{y}\end{array}\right.\\
\newline
\infty&{\rm elsewhere}\end{array}\right.
\end{equation}
With this definition of confinement potential (which provides the
required boundary conditions), we solve equation (2) numerically to
find eigenvalues and eigenfunctions of bound states of electron in
the presence of electric and magnetic fields. Our numerical strategy
based on finite difference algorithm is presented in the Appendix.
The results of these calculations for different external field
configurations and variety of possibilities are shown in Figures and
will be interpreted in section $4$. The bound states wave functions
obtained in this section will be used to study the bandgap
renormalization of H-shaped quantum wire in the next section.

\section{Many Body Effects in H-Shaped Quantum Wires}
There are several nonlinear optical many body effects in quasi-one
dimensional semiconductor systems originating from
exchange-correlation effects in a dense excited plasma. One of the
most important many-body effect in high density electron-hole plasma
is a density-dependent re-normalization of the fundamental band gap
of the semiconductor, which causes an increasing absorption in the
spectral region below the lowest exciton resonance. The
exchange-correlation correction of the fundamental band gap due to
the presence of free carriers (electrons in the conduction band and
holes in the valence band) in the system is referred to as the band
gap re-normalization effect. Optical nonlinearities, which are
strongly influenced by screened Coulomb interaction in the
electron-hole plasma, are typically associated with the band gap
re-normalization phenomenon. In which follows we use the two
band(one conduction band and one valance band) model to study the
one dimensional electron-hole system. We neglect the effects of
higher subbands and degeneracies in valance bands. We assume that
electrons and holes densities are constant in time. In this
situation, Hamiltonian of the system can be written as[8,9,33]
$$H=\sum_{k}\Bigg[\bigg(E^{0}_{g}+\frac{k^2}{2m_{e}}c^{\dag}_{k}c_{k}+
\frac{k^2}{2m_{h}}d^{\dag}_{k}d_{k}\bigg)\Bigg]$$
\begin{equation}
+\frac{1}{2L}\sum_{k,k',q}\bigg[V_{c,ee}(q)c^{\dag}_{k-q}c^{\dag}_{k'+q}c_{k'}c_{k}+
V_{c,hh}(q)d^{\dag}_{k-q}d^{\dag}_{k'+q}d_{k'}d_{k}+
V_{c,eh}(q)c^{\dag}_{k-q}c_{k}d^{\dag}_{k'+q}d_{k'}\bigg].
\end{equation}
In this equation which contains all information about this one
dimensional system, $c_{k}$ and $c^{\dag}_{k}$ are annihilation and
creation operators for conduction electrons respectively. Also
$d_{k}$ and $d^{\dag}_{k}$ are annihilation and creation operators
for valance band holes. $E^{0}_{g}$ is the direct band gap between
the top of the valance band and the bottom of the conduction band.
$V_{c,ij}$ show the possible three Coulomb interactions between
electrons and holes. Note that the two first interactions lead to
electron-hole quasi particle self-energies while the third one leads
to the production of excitonic bound states. Note also that this
Hamiltonian consists of spin effects, although spin index is not
included explicitly.

The Coulomb interaction matrix element in one dimensional quantum
wire is given by the following relation[8]
$$V_{c,ij}(q)=\frac{e^2}{\epsilon_{0}}\int_{-\infty}^{+\infty}dxdx'\int_{-\infty}^{+\infty}dydy'\int_{-\infty}^{+\infty}dz
\frac{e^{-iqz}|\phi_{i}(x,y)|^{2}
|\phi_{j}(x',y')|^{2}}{\sqrt{z^2+(y-y')^2+(x-x')^{2}}}$$
\begin{equation}
=\frac{2e^2}{\epsilon_{0}}\int_{-\infty}^{+\infty}dxdx'\int_{-\infty}^{+\infty}dydy'|\phi_{i}(x,y)|^{2}
|\phi_{j}(x',y')|^{2}K_{0}\Big(q\sqrt{(x-x')^2+(y-y')^{2}}\Big),
\end{equation}
where $\phi_{i}(x,y)$ is the quantum wire confinement wave function
for the lowest eigenstate of electrons or holes. The exact form of
these eignfunctions depends on the geometry and details of
confinement potential. $K_{0}(x)$ is the zeroth-order modified
Bessel function of second kind. In the setup of our one dimensional
quantum system we have assumed that carriers are free to move in $z$
direction but $x$ and $y$ are directions of confinement.

Band gap renormalization in quasi-static approximation is given
by[14,15]
\begin{equation}
\Delta_{i}(k)=\sum_{k'}\Big[-V_{s}(k-k')n_{i}(\varepsilon_{i,k})+\frac{1}{2}(V_{s}(k')-V_{c}(k'))\Big]
\end{equation}
where
\begin{equation}
V_{s}(k)\equiv
V_{s}(k,\omega=0)=\frac{V_{c}(k)}{\varepsilon(k,\omega=0)}
\end{equation}
is the statically screened Coulomb interaction and
$n_{i}(\varepsilon_{i,k})$ is the fermion momentum distribution
function
\begin{equation}
n_{i}(\varepsilon_{i,k})\equiv
(e^{\beta(\varepsilon_{i,k}-\mu_{i})}+1)^{-1}
\end{equation}
where $ \varepsilon_{e,k}\equiv\frac{k^2}{2m_e}+E^{0}_{g}$ and $
\varepsilon_{h,k}\equiv\frac{k^2}{2m_h}$ are the bare energies of
electron and hole in their respective bands and $\mu_{i}$ is the
chemical potential. $\varepsilon(k,\omega)$ as dynamical dielectric
function is defined as follows
\begin{equation}
\varepsilon(k,\omega)=1-V_{c}(k)\Pi^{0}_{e}(k,\omega)-V_{c}(k)\Pi^{0}_{h}(k,\omega)=
1-V_{c}(k)\sum_{i=e,h}\frac{m_{i}}{\pi
k}\ln\bigg[\frac{\omega^{2}-[(k^{2}/2m_{i})-k\upsilon_{F,i}]^{2}}
{\omega^{2}-[(k^{2}/2m_{i})+k\upsilon_{F,i}]^{2}}\bigg]
\end{equation}
where $\upsilon_{F,e/h}$ is the Fermi velocity of electrons/holes at
Fermi momentum in the conduction/valance band.

In one-loop GW approximation with dynamically screened interaction,
one has[8,33]
\begin{equation}
\Sigma_{i}(k,z)=-\frac{1}{\beta}\sum_{k',z'}V_{s}(k-k',z-z')G_{i}(k',z')=
-\frac{1}{\beta}\sum_{k',z'}\frac{V_{c}(k-k')}{\varepsilon(k-k',z-z')}G_{i}(k',z')
\end{equation}
where
\begin{equation}
G_{i}(k,z)=\frac{1}{z-\varepsilon_{i,k}-\Sigma_{i}(k,z)+\mu_{i}},\quad\quad\quad
(i=e,h)
\end{equation}
and  $\Sigma_{e/h}(k,z)$ is the self energy of electrons/holes
defined in GW approximation. To avoid multi-pole structure in
$G_{i}(k,z)$ we approximate $\Sigma_{i}(k,z)$ by momentum-dependent
band gap renormalization $\Delta_{i}(k)$. Using the approximation
$\Delta_{i}(k)= \Sigma_{i}(k,\varepsilon_{i,k}-\mu_{i})$, we find
the following single pole electron-hole Green's function[8,33]
\begin{equation}
G_{i}(k,z)\sim\frac{1}{z-\varepsilon_{i,k}-\Delta_{i}(k)
+\mu_{i}},\quad\quad\quad(i=e,h)
\end{equation}
The above formalism provides a suitable framework for our numerical
calculation of band gap renormalization. To proceed further, we
should calculate the screened coulomb potential. Using equation (5),
it can be written as follows
$$V_{c}(k)=\frac{2e^2}{\varepsilon_{0}}\int dx
dy\int dx'dy' K_{0}|k R| |\Psi(x,y)|^2
 |\Psi(x',y')|^2,$$

or using rescaled coordinates $\tilde{x_{i}}\equiv \frac{x_{i}}{L}$
and  $\tilde{x_{i}}'\equiv \frac{x_{i}'}{L}$ we find
\begin{eqnarray}
V_{c}(k)=\lambda\frac{2e^2}{\varepsilon_{0}} L^4\int d\tilde{x}
d\tilde{y}\int d\tilde{x}'d\tilde{y}' K_{0}|k R|
|\Psi(\tilde{x},\tilde{y})|^2
 |\Psi(\tilde{x}',\tilde{y}')|^2,
\end{eqnarray}
where $\lambda$ is re-scaling factor equal to $10^{-18}/m^2$ and $k
R =k\sqrt{(x-x_{0})^2+(y-y_{0})^2}$. Using $\Psi(x,y)$ computed
numerically in the previous section, we solve the integral of
equation (13). The screened potential is calculated by ground state
wave function as a function of k. Figure 15 shows the result of this
calculation for different values of relative width of the
confinement potential. In this figure, $V_{c}(k)$ is normalized by
$\frac{2e^2}{\epsilon_0}$ and the $k$ is normalized to $kL$.\\
In the next step we calculate the band gap renormalization in both
quasi static and GW approximation. To do this end, we should
calculate some quantities numerically. Using the re-scaled
quantities $\tilde{k}=kL$ and $\omega=\tilde{\omega}\times10^{16}$,
we write relation (9) as follows
\begin{eqnarray}
\varepsilon(k,\omega)=1-V_{c}(k')
\Bigg(\frac{1.7\times10^{45}}{\tilde{k}L}\ln\Bigg[\frac{\tilde{\omega}^2\times10^7-525[12\frac{\tilde{k}^2}{L^2}-\frac
{\tilde{k}}{L}]^2}{\tilde{\omega}^2\times10^7-525[12\frac{\tilde{k}^2}{L^2}+\frac{\tilde{k}}{L}]^2}\Bigg]\\
+\frac{5.18\times10^{44}}{\tilde{k}L}\ln\Bigg[\frac{\tilde{\omega}^2\times10^6-174[43.8\frac{\tilde{k}^2}{L^2}-\frac
{\tilde{k}}{L}]^2}{\tilde{\omega}^2\times10^6-174[43.8\frac{\tilde{k}^2}{L^2}+\frac{\tilde{k}}{L}]^2}\Bigg]\Bigg).
\end{eqnarray}
The single pole Green's function for electron defined as relation
(12) transforms to the following form
\begin{equation}
G(k,z)=\frac{1}{1.06\times10^{-26}i\tilde{\omega}-6.17\times10^{-21}\frac{\tilde{k}^2}{L^2}-2.47\times10^{-19}-\Delta_{e}(k)}
\end{equation}
and Green's function for hole becomes
\begin{equation}
G(k,z)=\frac{1}{1.06\times10^{-26}i\tilde{\omega}-9.2\times10^{-20}\frac{\tilde{k}^2}{L^2}+1.6\times10^{-22}-\Delta_{h}(k)}.
\end{equation}
To calculate bandgap renormalization in this configuration, we
define the re-scaled $\beta_e$ and $\mu_e^0$ respectively as
$\frac{\beta \hbar^2}{2m_e^* W_y^2}=\beta \frac{574.5}{W_y^2}$ and
$\mu^0_e \beta$, where $W_y$ is width of quantum well in $y$
direction in Nanometer. For holes, we also define the re-scaled
$\beta_h$ and $\mu^0_h$ as $\beta_h =\beta_e
\big(\frac{m^*_e}{m^*_h} \big)$ and $\mu_h^0=\mu_e^0
\big(\frac{m^*_e}{m^*_h} \big)$ respectively. In all computations in
this paper, we have assumed that the ratio $\frac{m_h^*}{m_e^*}$ is
equal to $0.3$ and $m_e^*\simeq 0.067 m_e$. Then we set
$\tilde{\omega}=0$ and using equation (7), we calculate band gap
renormalization in both quasi static and GW approximations at
temperature $T=0$. The result of calculation for both quasi static
and GW approximation is shown in figure 16. As this figure shows,
the value of band gap renormalization in GW approximation is smaller
than the quasi static plasmon-pole approximation. Typical values of
band gap normalization are between 10-30 $meV$ depending on the
temperature and impurities in the system. Because of consideration
of more quantum field theory details, GW approximation generally
gives results which have better agreement with experimental
results[10,11].

\section{Interpretation of Numerical Results and conclusions}
Probability distribution of charge carriers in the absence of
external electric and magnetic fields are shown in figure $2$ where
the asymmetric parameter $a=\frac{W_{x}}{W_{y}}$ has been set equal
to $0.8$. This figure emphasizes the central role played by
geometric shape of the confinement potential. In another words, in
the absence of external electric and magnetic fields, carrier
distribution obeys the symmetry of confinement potential. Variation
of asymmetric parameter changes the profile of charge distribution
in such a way that the case with $a=1$ has maximum symmetry and any
change of relative width leads to antisymmetric distribution. Figure
$3$ shows the variation of the ground state energy of carrier versus
the inverse of the asymmetric parameter $a$. Variation of the
relative width leads to the conclusion that smaller relative width
leads to smaller ground state energy. Now, suppose that we turn on a
uniform magnetic field. In the absence of electric field the
distribution of charge carriers is given by figure $4$. The role
played by asymmetric parameter is a reduction of probability
amplitude when the width of the well increases in $x$ or $y$
direction. Variation of ground state energy versus the intensity of
magnetic field is depicted in figure $5$ for two different values of
asymmetric parameter. For a fixed well width in $y$ direction, when
the width of the well in $x$ direction increases the value of the
ground state energy will increase.\\
Now we turn off magnetic field and apply a uniform electric field in
the $x$ direction. The profile of probability amplitude for charge
carriers distribution is shown in figure $6$ for $a=0.8$. The
probability amplitude has a Gaussian profile and is shifted toward
the right hand side. This shift is a function of electric field
intensity. The probability profile does not obey the geometric shape
of the confinement potential due to preferred direction defined by
the presence of electric field. The energy of the ground state
versus the intensity of electric field is depicted in figure $7$. As
this figure shows, ground state energy decreases with increasing
electric field intensity. This is not surprising since external
electric field tends to decouple electrons and holes from each
other. In the general case where both electric and magnetic fields
are present, the asymmetric situation explained above will be
enhanced in some respects. Figures $8$ shows the space variation of
the probability amplitude with $a=0.8$. In the presence of constant
electric and magnetic fields, pick of the graph describing charge
carriers distribution will be shifted to the positive direction of
the $x$ axis. This feature causes the carriers concentration in such
a way that these carriers distribution do not obeys the external
confinement potential symmetry. More the intensity of the electric
field results in more shift of the distribution pick to the right
hand side of the $x$ axis. The presence of constant magnetic field
causes the anisotropy in the profile of the probability distribution
of the carriers. Figure $9$ shows the variation of the ground state
energy versus the variation of the electric field when the magnetic
field is supposed to be constant. More intensity of the electric
field leads to the more reduction of the bound states energies. This
resembles the linear Stark effect in elementary quantum mechanics.
In the language of many body effects in dense plasma in a quasi one
dimensional semiconductor, application of intense electric field
results in the weaker excitonic states. Therefore, the presence of
external electric and magnetic fields will shift the location of
maximum concentration of carriers and in this case there is an
apparent asymmetry in profile of carriers distribution. This point
can be used in fabrication of microelectronic devices based on
quantum wires.

We have proposed a numerical framework to calculate screened Coulomb
potential and the values of band gap energy re-normalization in
H-shaped quantum wire in two different approximations: quasi-static
and dynamical random phase approximation in its leading order
dynamical screening (GW approximation). We have evaluated the single
particle self-energies for both electrons and holes in the dynamical
plasmon pole approximation (PPA) and the leading order dynamically
screened interaction or GW approximation to obtain the electron and
hole renormalized Green's functions. This self-energy calculation
gives us the band gap renormalization due to exchange-correlation
effect. For comparison, we also calculated the band gap
renormalization obtained by the quasi static calculation in both
static random phase approximation and static plasmon pole
approximation. Note that quasi static approximation works well in
two and three dimensional systems but fails completely in one
dimensional systems. This is because the electrons in one
dimensional system suffer very strong inelastic scattering effects
by virtue of restricted phase space. However, we have used this
approximation only for comparison purposes. It is important to note
that by re-scaling procedure which we have considered, we have fixed
the geometry of the H-shaped quantum wire. Actually, one should
consider the possibility for changing the geometry also. This has
been down by change in parameter $L$ but our calculations show that
the main physical results do not change considerably.

Figure 10 shows the screened Coulomb potential calculated based on
random phase approximation in its leading order and for different
values of asymmetry parameter. Exchange-correlation many body
effects mediate the bare Coulomb interaction. Based on different
width of the confinement potential, screened Coulomb potential
varies with geometrical characteristics of confinement potential. As
the figure 10 shows, by increasing the asymmetry parameter, screened
Coulomb potential will grows but its overall behaviors with respect
to wave number will not change. Figure $10$ shows the calculated
band gap renormalization  in quasi static and dynamical random phase
approximations. To calculate band gap renormalization we first
calculate the electron/hole single pole Green's function and then
using the formalism of both one-loop GW approximation and quasi
static plasmon pole approximation we calculate the values of band
gap renormalization in $T=0$. Note that we have considered $L$
dependence of band gap renormalization, however it can be translated
to band gap renormalization versus carrier densities as well. Figure
11 also compares the results of band gap renormalization in GW and
plasmon-pole approximations. As this figure shows, for a fixed value
of $L$, GW approximation gives smaller values of band gap
renormalization. Generally GW approximation gives more reliable
result in comparison with experimental data[5]. Note that we have
focused on the variation of geometry by variation of $L$. As figure
1 shows there are other possibilities for changing the geometrical
shape of external confinement potential. However the general
behavior is the same as presented here. There are some restrictions
on our calculations which can be summarized as follows: the
many-body treatment has the disadvantage that, for band gap
renormalization it commonly ignores geometrical factors, such as the
quantum confined Stark effect, whose relevance is structure
specific. In other words, the numerical results for different
confinement potential of H-shaped quantum wire may be geometry
dependent in general. Furthermore, in the exciton problem, many-body
theory treats screening within the linear approximation and,
generally, influence of the bound electron on the free electrons is
not fully included. In particular the orthogonality of the free
electron states with the bound state, which increases its importance
in lower dimensional systems, are typically not included. On the
other hand the complete treatment of the problem should consider the
effects of several subbands[8]. Note that in the rest of the
calculation of band gap renormalization we have used the two
band(one conduction band and one valance band) model and we have
neglected the effects of higher subbands and degeneracies in valance
bands. We also assumed that electrons and holes densities are
constant in time. In summary, many body effects in quasi one
dimensional systems are sensitive to the geometrical shape of
external confinement effects. The value of band gap renormalization
in GW approximation is smaller than quasi static plasmon-pole
approximation. Typical values of this normalization are between
10-30 $meV$ depending to temperature and impurities in the system.
GW approximation generally gives results which have better agreement
with experiment. A complete study of this problem requires the
considerations of several subbands and varying electrons and holes
densities.

\section{Summary and Conclusions}
Our numerical procedure to the bound states and many-body effects in
H-shaped quantum wires has the following results:
\begin{itemize}
\item
The distribution of the probability of the charge carriers in the
ground state of the H-shaped quantum wire in the absence of electric
and magnetic fields has a symmetric shape obeying the geometric
shape of the confinement potential. This distribution has its
maximum at the center of each arm and decreases with distance from
the center. The relative width of the confinement potential(the
asymmetric ratio $a=\frac{W_{x}}{W_{y}}$ has considerable effect on
the profile of this distribution.
\item
In the presence of a constant magnetic field, the distribution of
the charge carriers becomes oscillatory both in $x$ and $y$
directions. Increasing the strength of the magnetic field leads to
the reduction of the ground state energy. The role played by the
asymmetric parameter is given by the reduction of the probability
amplitude when the width of the well increases in $x$ or $y$
directions.
\item
The situation for the case of non-vanishing electric field (in the
absence of magnetic field) resembles the {\it Stark effect} in a low
dimensional system. The probability amplitude has an oscillatory
behavior with larger wavelength of the oscillations. In this case
although the probability amplitude has a Gaussian profile, it is
shifted toward the right hand side. This shift is a function of the
electric field intensity. The probability profile does not obey the
symmetry of the geometric shape of the confinement potential due to
preferred direction defined by the presence of the electric field.

\item
In the presence of both electric and magnetic fields there are
oscillations in probability distribution both in $x$ and $y$
directions, but in this case the probability distribution is not
symmetric. In the presence of constant electric and magnetic fields,
the pick of the probability amplitude will be shifted along the
positive direction of the $x$ axis. This causes the carriers to be
concentrated in such a way that they do not obey the external
confinement potential symmetry. More intensity of the electric field
results in more shift of the distribution pick to the right hand
side of the $x$ axis. The presence of constant magnetic field causes
an anisotropy in the profile of the probability distribution of the
carriers since it apparently breaks the local rotational symmetry in
the center of each arm.

\item
Screened Coulomb potential of the H-shaped external confinement is a
sensitive function of the asymmetry parameter but its general
behavior under variation of wave number is the same for other
possible geometries of quasi one dimensional systems. The calculated
band gap narrowing for H-shaped confinement potential in the absence
of the electric and magnetic fields and within quasi-static and
dynamical random phase approximation shows a typical gap narrowing
of the order of few $meV$. This is supported from other studies of
band gap narrowing in quasi one dimensional semiconductor
systems[4,6]. The dynamical random phase approximation leads to more
reliable result than quasi static approximation in comparison with
experiments.
\end{itemize}

In summary we can conclude that in the presence of electric and
magnetic fields, H-shaped quantum wires bound states characteristics
are sensitive functions of an asymmetric parameter
$a=\frac{W_{x}}{W_{y}}$ and the strength of the electric and
magnetic fields. The case of non-vanishing electric and magnetic
fields induces an intrinsic inhomogeneity in the quasi one
dimensional system. Many-body effects due to plasma screening and
resulting optical nonlinearities are also dependent to the
asymmetric parameter of quasi-one dimensional confinement potential.
Among these nonlinear optical effects, band gap renormalization of
fundamental band age has been studied numerically in this paper.\\

\begin{center}
\Large{\bf Appendix: Numerical Strategy}
\end{center}
\small \vspace{1cm}

We use the finite difference algorithm[34] to solve our partial
differential equation(2) with boundary conditions imposed by
H-shaped confinement potential. The most straightforward refinement
method replaces the differential equation with a finite difference
equation. It replaces all derivatives with approximate expression as
the following familiar form
$$\Bigg(\frac{dy}{dx}\Bigg)_{x=x_{n}}\approx
\frac{y_{n+1}-y_{n}}{h}$$
\begin{equation}
\Bigg(\frac{d^{2}y}{dx^{2}}\Bigg)_{x=x_{n}}\approx
\frac{y_{n+1}-2y_{n}+y_{n-1}}{h^2}
\end{equation}
where the mesh chosen to be equally spaced and given by \,
$h=\frac{x_{k}-x_{0}}{N+1}$\, with $x_{k}=x_{0}+kh$ and $k=1,\,2,\,
...,n+1$. We first re-scale the Schr\"{o}dinger equation and the
confinement potential. We define the re-scaled value of $x$ and $y$
as $\bar{x}\equiv \frac{x}{W_{x}}$ and
$\bar{y}\equiv\frac{y}{W_{y}}$ where $W_{x}$ and $W_{y}$ are the
width of the well in $x$ and $y$ direction respectively. Now the
re-scaled Schr\"{o}dinger equation can be written as follows
$$-\Bigg(\frac{1}{W_{x}}\frac{\partial^{2}}{\partial
\bar{x}^{2}}+\frac{1}{W_{y}}\frac{\partial^{2}}{\partial
\bar{y}^{2}}-\frac{2ieBW_{x}}{\hbar
W_{y}}\bar{x}\frac{\partial}{\partial
\bar{y}}\Bigg)\Psi(\bar{x},\bar{y})+
\Bigg(\frac{eBW_{x}}{\hbar}\Bigg)^{2}\bar{x}^{2}\Psi(\bar{x},\bar{y})$$
\begin{equation}
- \frac{2meEW_{x}}{\hbar^{2}}\bar{x}\Psi(\bar{x},\bar{y})
=\frac{2m}{\hbar^{2}}\lambda\Psi(\bar{x},\bar{y}).
\end{equation}
Using the finite difference algorithm, this equation can be written
as follows
$$-\Bigg(\frac{\Psi_{i+1,j}-2\Psi_{i,j}+\Psi_{i-1,j}}{\Delta_{x}}+
a^{2}\frac{\Psi_{i,j+1}-2\Psi_{i,j}+\Psi_{i,j-1}}{\Delta_{y}}\Bigg)$$
\begin{equation}
+\frac{2ieaBW_{x}^{2}}{\hbar}\bar{x_{i}}\frac{\Psi_{i,j+1}-\Psi_{i,j-1}}{2\Delta_{y}}
+\Bigg(\frac{eBW_{x}^{2}}{\hbar}\Bigg)^{2}\bar{x_{i}}^{2}\Psi_{i,j}-
\frac{2meEW_{x}^{3}}{\hbar^{2}}\bar{x_{i}}\Psi_{i,j}
=\frac{2mW_{x}^{2}}{\hbar^{2}}\lambda\Psi_{i,j}.
\end{equation}
where $a=\frac{W_{x}}{W_{y}}$ is the relative width of the H-shaped
quantum wire. We do discritize the $x$\, and\,  $y$ axes to discrete
space $dx=0.1$ \, and \, $dy=0.1$, therefore the equation (20) can
be written as a matrix equation
\begin{eqnarray}
 {\cal H} \phi = E \phi,
\end{eqnarray}
where  $\cal H$, and $\phi$ are the Hamiltonian matrix and the state
wave functions array which are defined as follows
\begin{eqnarray}
 \phi =
 \left(\begin{array}{c}
\left(\vdots\right) \\
\left(\begin{array}{c}
\vdots \\
\phi(x_i,y_j) \\
\phi(x_{i+1},y_j) \\
\vdots
\end{array}
\right ) \\
\left(\begin{array}{c}
\vdots \\
\phi(x_i,y_{j+1}) \\
\phi(x_{i+1},y_{j+1}) \\
\vdots
\end{array}
\right ) \\
\left(\vdots\right)
\end{array}
\right )
 \end{eqnarray}
\begin{tiny}
\begin{eqnarray}
{\cal H}=\left(\begin{array}{cccc}
  \left(\ddots \right) & \left(\begin{array}{ccc}
  \ddots & 0 & 0   \\
  0 & -\frac{a^{2}}{\Delta_{y}^{2}}+\frac{iW_{x}eaB\bar{x_{i}}}{\hbar}& 0  \\
  0 & 0 & \ddots
  \end{array}\right) & 0 & 0  \\
  \left(\ddots \right) &
  \left(\begin{array}{cccc}
  \ddots & \ddots & 0 & 0  \\
  \ddots &2(\frac{1}{\Delta_{x}^2}+\frac{a^{2}}{\Delta_{y}^2})+
  (\frac{eW_{x}^{2}B}{\hbar})^{2}\bar{x_{i}}^{2}+\frac{2meW_{x}^{3}E}{\hbar^{2}}
  \bar{x_{i}} & \frac{-1}{\Delta_{x}^2} & 0 \\
  0 & \frac{-1}{\Delta_{x}^2}
  &2(\frac{1}{\Delta_{x}^2}+\frac{a^{2}}{\Delta_{y}^2})+
  (\frac{eW_{x}^{2}B}{\hbar})^{2}\bar{x_{i}}^{2}+\frac{2meW_{x}^{3}E}{\hbar^{2}}
  \bar{x_{i}}  & \ddots \\
  0 & 0 & \ddots & \ddots
  \end{array}\right)
  & \left(\ddots\right)
  & 0   \\
  0 &
\left(\begin{array}{ccc}
  \ddots & 0 & 0   \\
  0 &  -\frac{a^{2}}{\Delta_{y}^{2}}-\frac{iW_{x}eaB\bar{x_{i}}}{\hbar}& 0  \\
  0 & 0 & \ddots
  \end{array}\right)
    & \left(\ddots\right)& \left(\ddots\right) \\
  0 & 0 & \left(\ddots \right)& \left(\ddots\right)
\end{array} \right )
\end{eqnarray}
\end{tiny}
The error of computation of $\phi(x,y)$ is of the order of
$O(dx^2)$. We diagonalize the Hamiltonian matrix and calculate the
bound states energies and wave functions numerically. We use MATLAB
package since it uses techniques that are more efficient than Jacobi
rotations and that can be applied to asymmetric or even complex
matrices as well as to the more common real symmetric situations.

\newpage

\begin{figure}[ht]
\begin{center}
\includegraphics{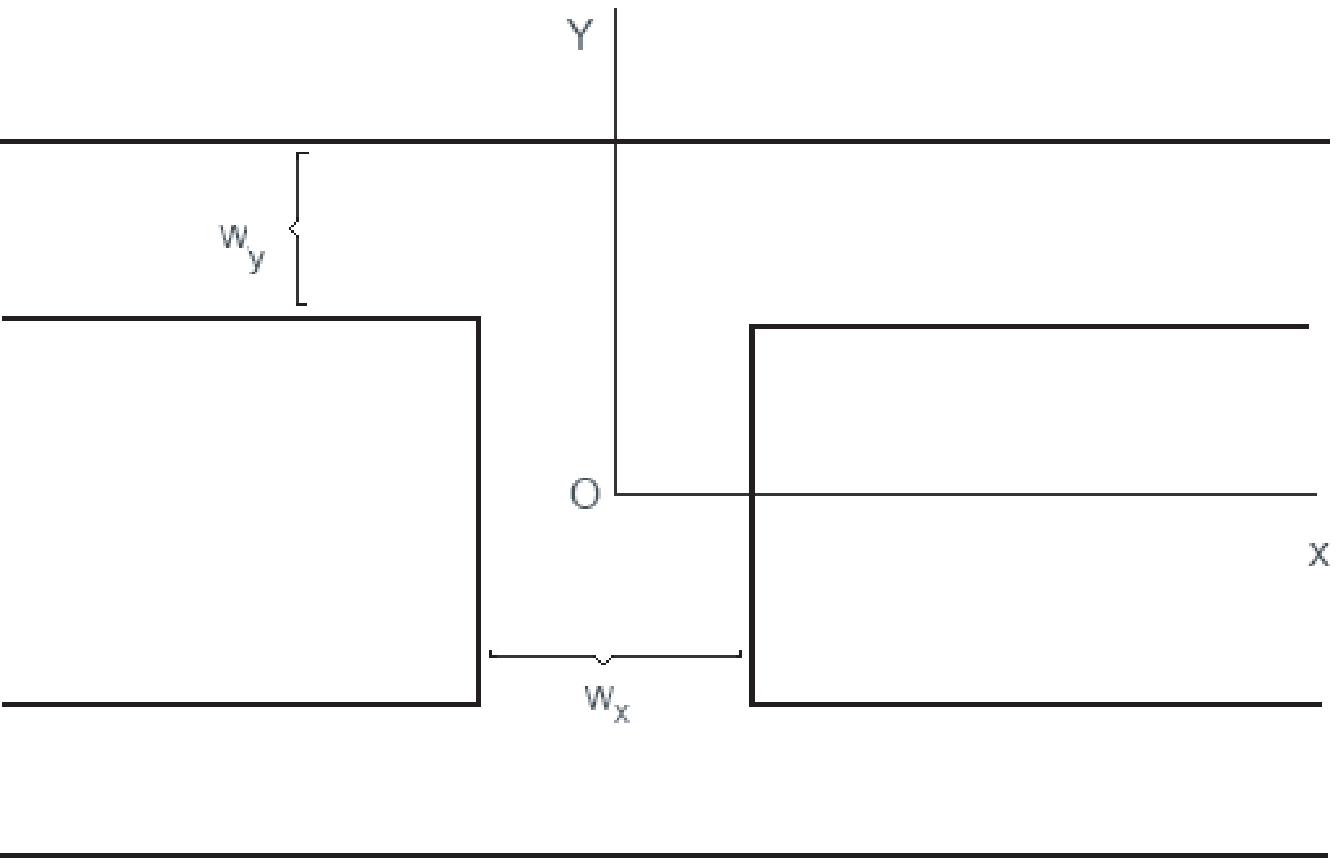}

\end{center}
\vspace{10 cm}
 \caption{\small {The Geometry of the H-Shaped Quantum Wire }}
 \label{Fig:2}
\end{figure}

\begin{figure}[ht]
\begin{center}
\includegraphics{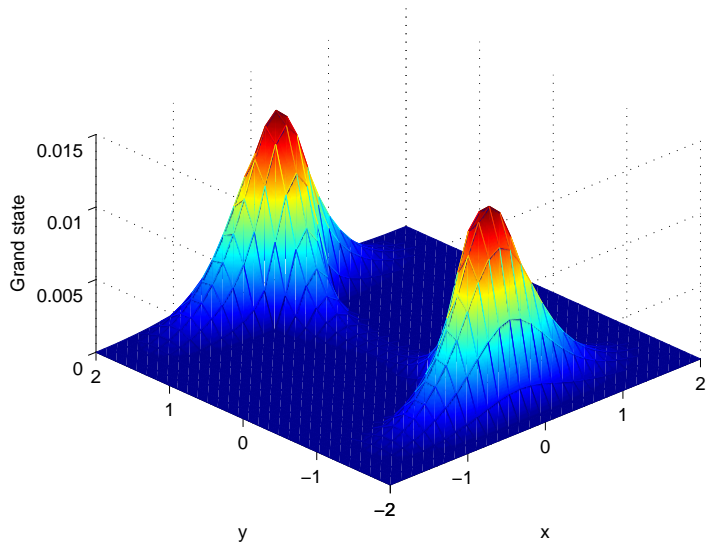}

\end{center}
\vspace{10 cm}
 \caption{\small {The probability amplitude of charge carriers
 distribution for ground state of H-shaped quantum wire at zero
 external EM fields. The asymmetry parameter has been chosen to
 be $a=\frac{W_{x}}{W_{y}}=0.8$.  }}
 \label{Fig:3}
\end{figure}

\begin{figure}[ht]
\begin{center}
\includegraphics{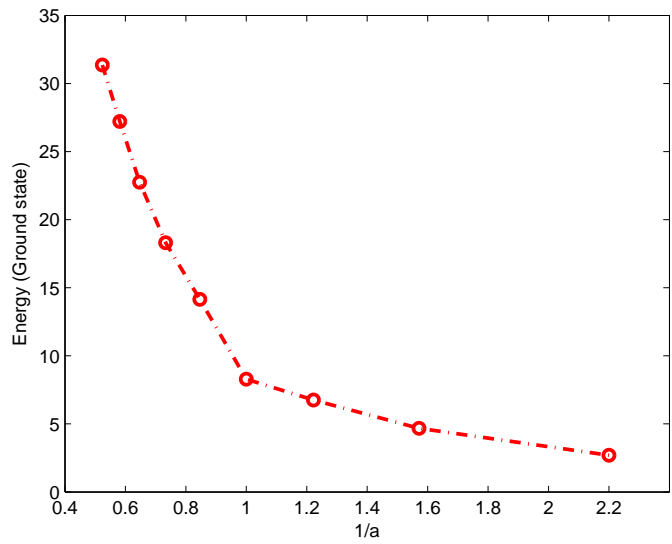}

\end{center}
\vspace{10 cm}
 \caption{\small {The ground state energy versus the inverse of the
 asymmetry parameter at zero external fields strength. }}
 \label{Fig:5}
\end{figure}

\begin{figure}[ht]
\begin{center}
\includegraphics{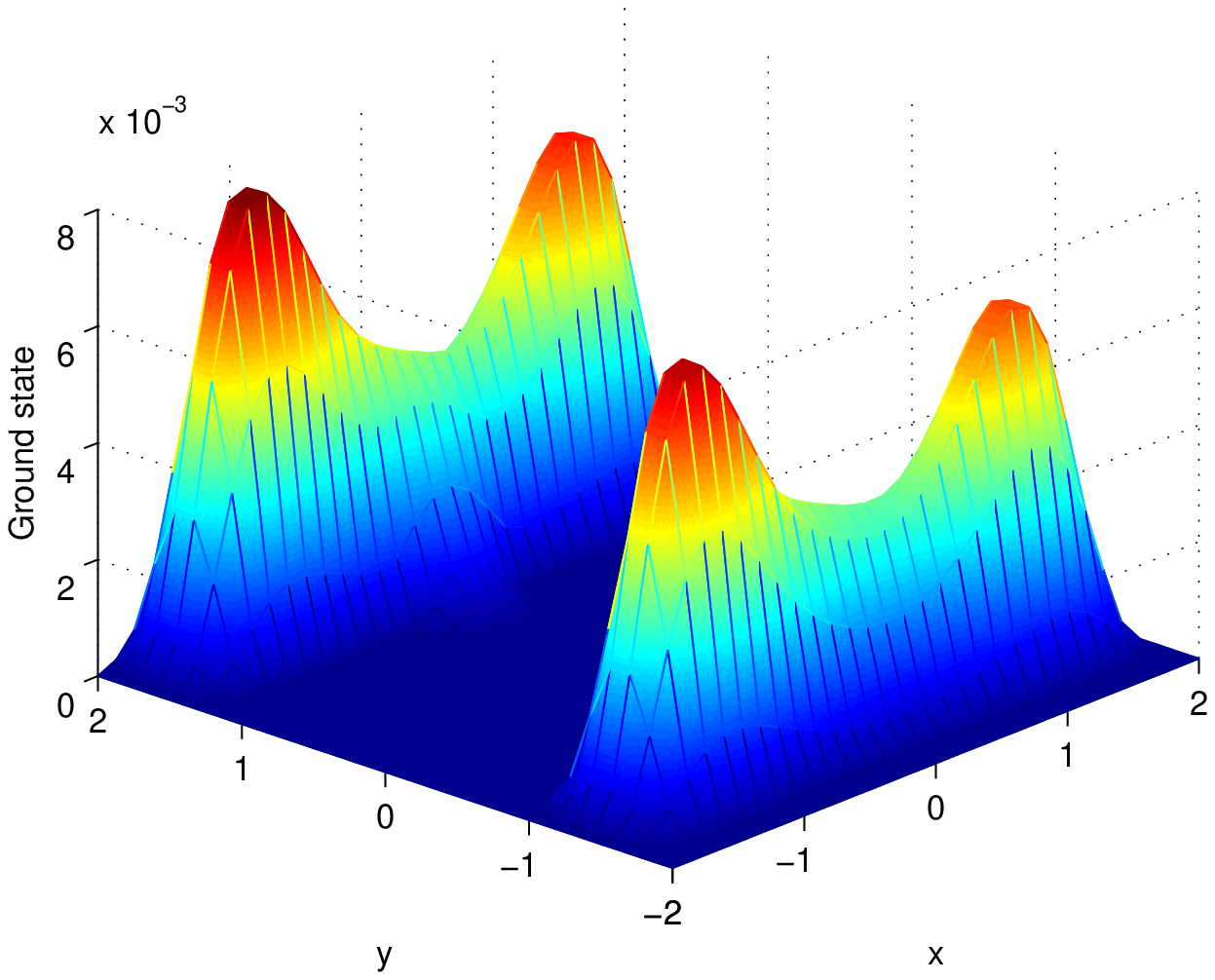}

\end{center}
\vspace{10 cm}
 \caption{\small {The probability amplitude of charge carriers
 distribution for ground state of H-shaped quantum wire in the presence
 of an external magnetic field(and zero electric field). The asymmetry
 parameter has been chosen to
 be $a=\frac{W_{x}}{W_{y}}=0.8$.}}
 \label{Fig:6}
\end{figure}

\begin{figure}[ht]
\begin{center}
\includegraphics{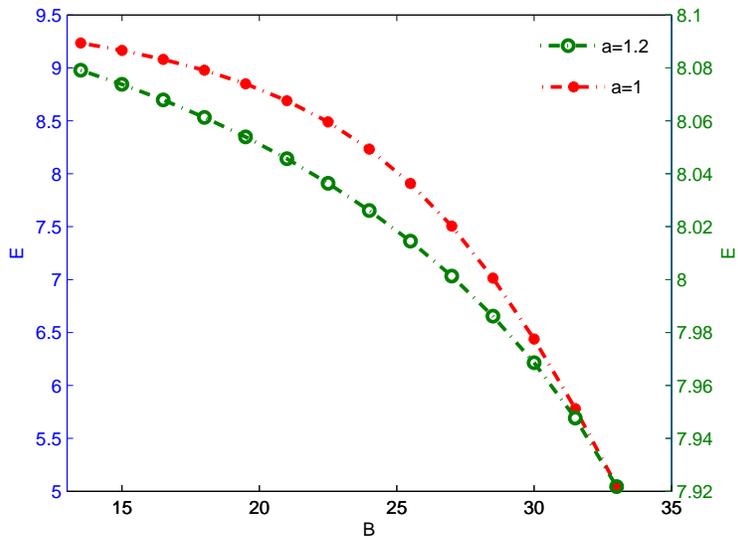}

\end{center}
\vspace{10 cm}
 \caption{\small {The ground state energy versus the intensity
 of the external magnetic field for different values of asymmetry parameter.}}
 \label{Fig:8}
\end{figure}

\begin{figure}[ht]
\begin{center}
\includegraphics{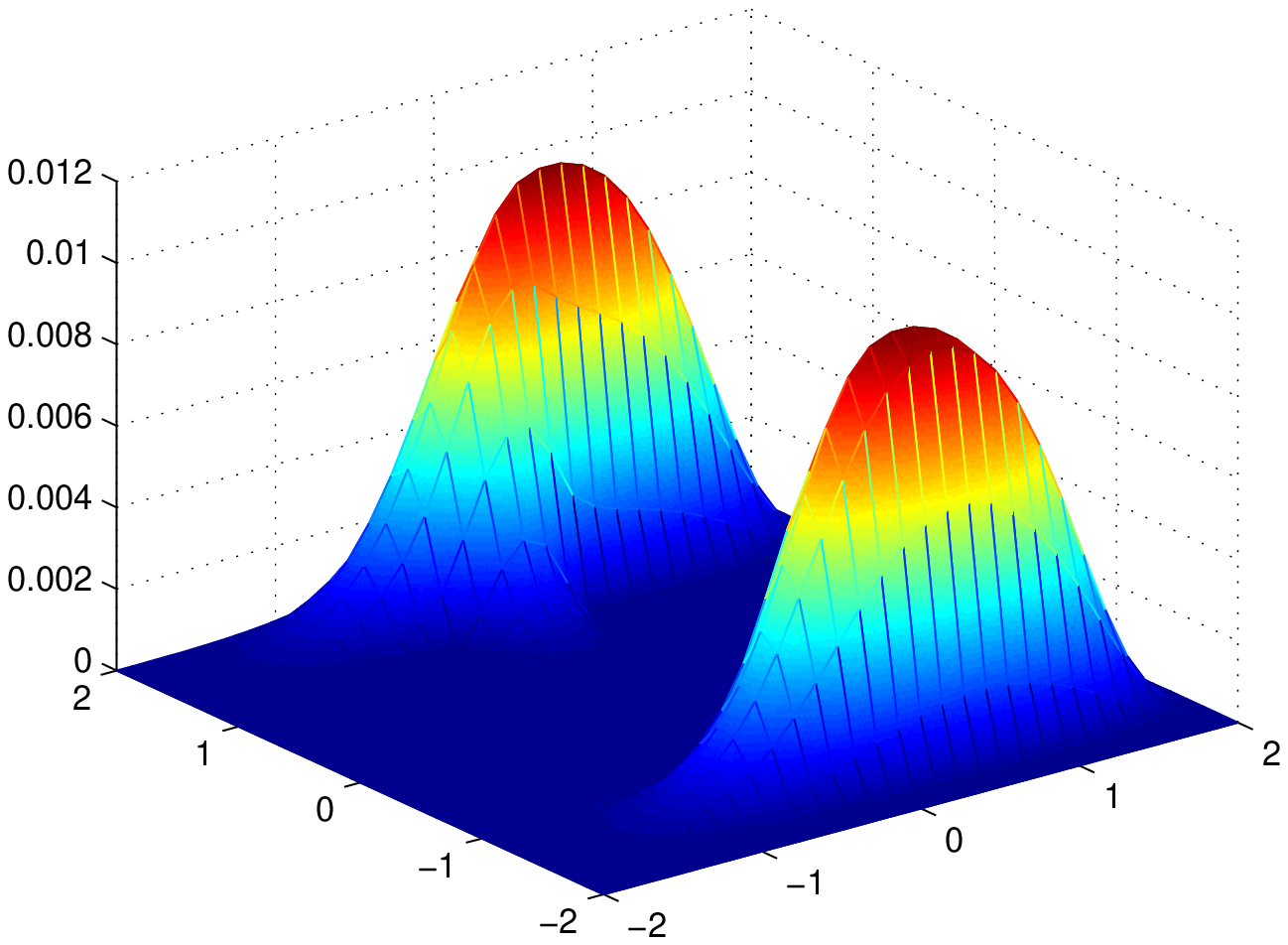}

\end{center}
\vspace{10 cm}
 \caption{\small {The probability amplitude of charge carriers
 distribution for ground state of H-shaped quantum wire in the presence
 of an external electric field(and zero magnetic field). The asymmetry
 parameter has been chosen to  be $a=0.8$.}}
 \label{Fig:10}
\end{figure}

\begin{figure}[ht]
\begin{center}
\includegraphics{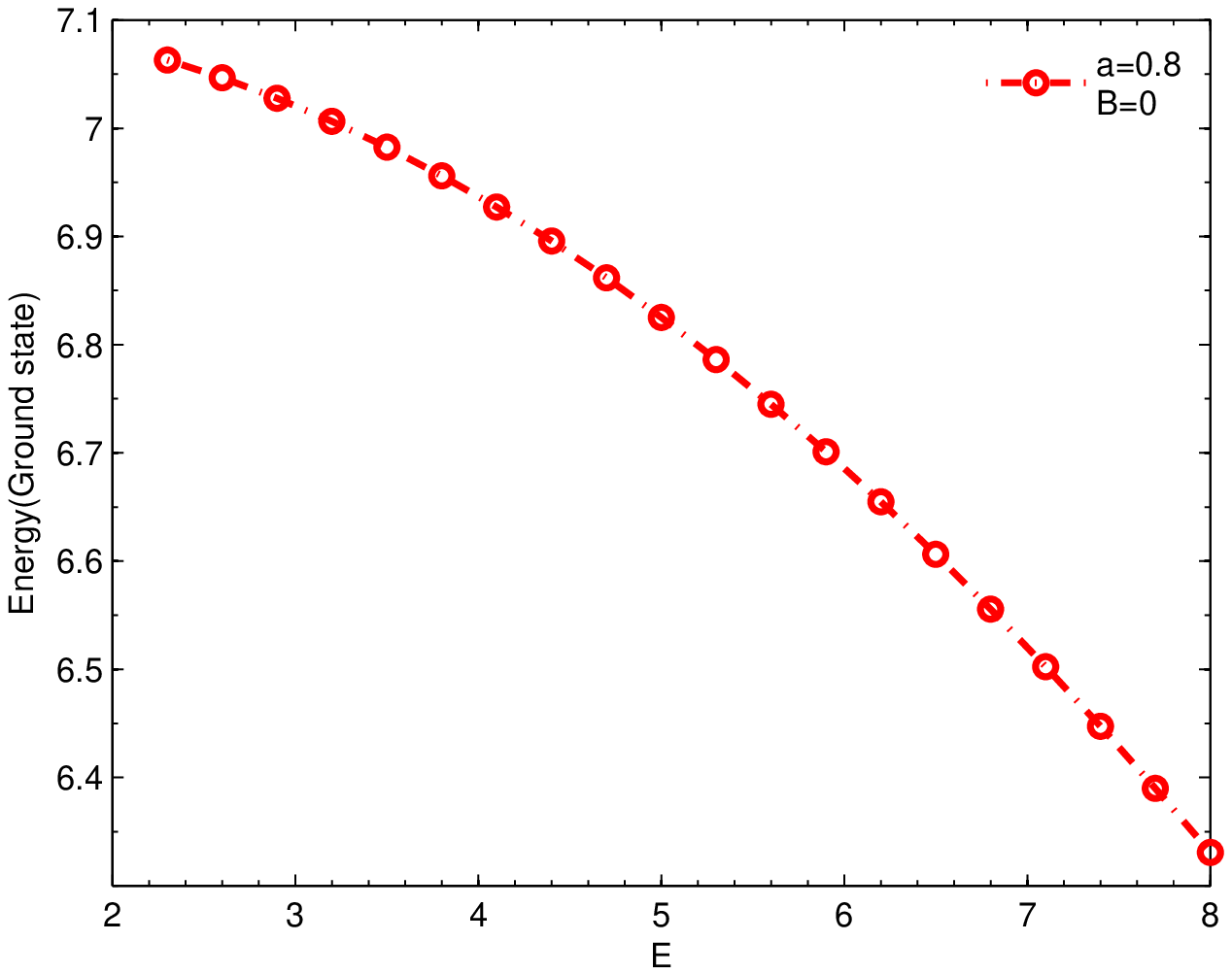}

\end{center}
\vspace{10 cm}
 \caption{\small {The ground state energy versus the intensity
 of the external electric field for $a=0.8$ and $B=0$.}}
 \label{Fig:12}
\end{figure}

\begin{figure}[ht]
\begin{center}
\includegraphics{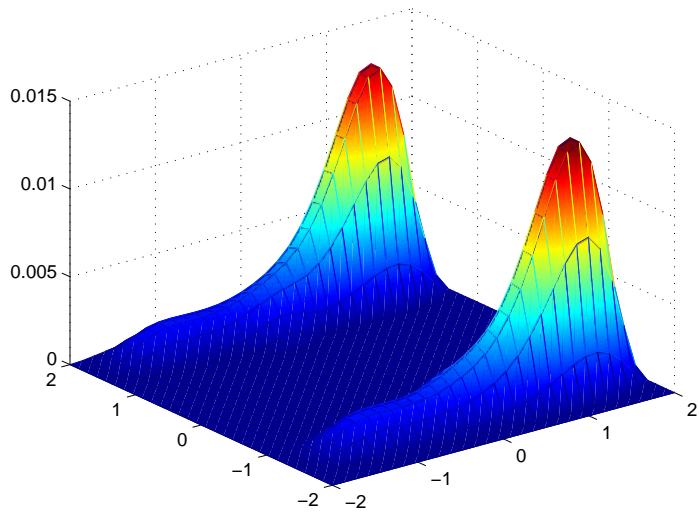}

\end{center}
\vspace{10 cm}
 \caption{\small {The probability amplitude of charge carriers
 distribution for ground state of H-shaped quantum wire in the
 presence of both electric and magnetic fields. The asymmetry
 parameter has been chosen to  be $a=0.8$. }}
 \label{Fig:13}
\end{figure}

\begin{figure}[ht]
\begin{center}
\includegraphics{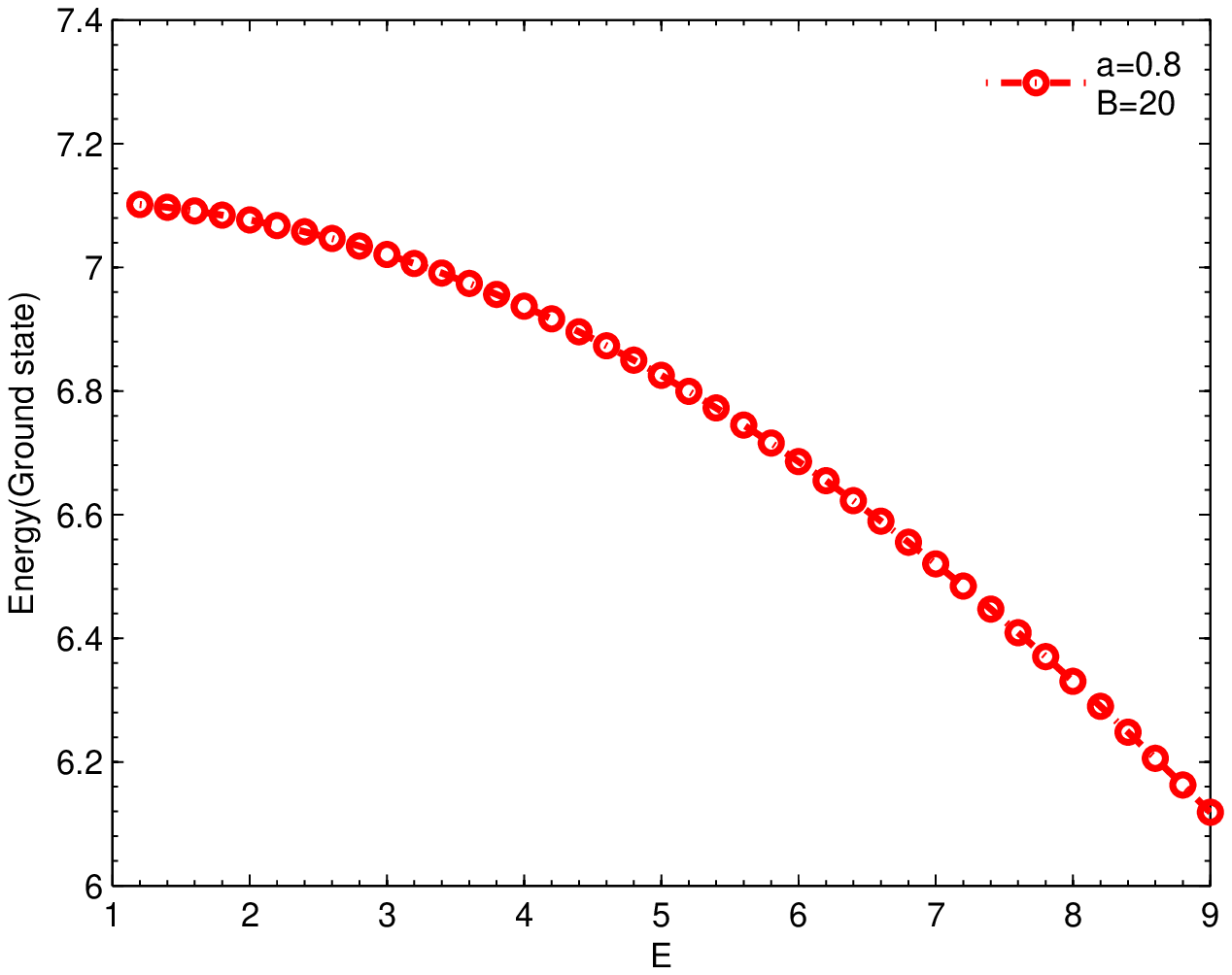}

\end{center}
\vspace{10 cm}
 \caption{\small { The ground state energy versus the intensity
 of the external electric field for $a=0.8$ and $B=20$ (a constant magnetic field).}}
 \label{Fig:15}
\end{figure}

\begin{figure}[ht]
\begin{center}
\includegraphics{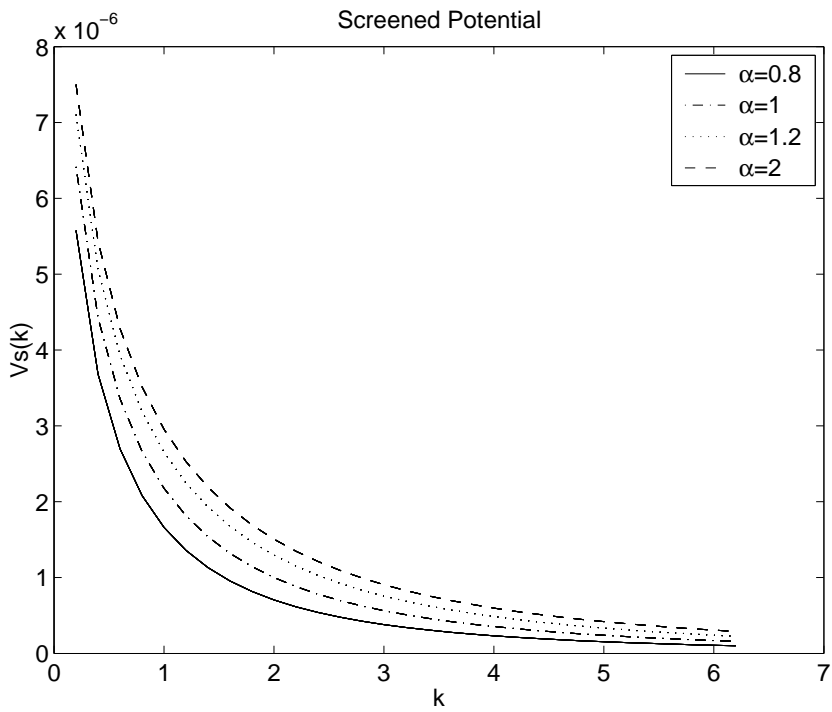}

\end{center}
\vspace{15 cm}
 \caption{\small {The calculated screened Coulomb potential
 versus the wave number for different asymmetry parameter.
 The screened potential is normalized by
$\frac{2e^2}{\epsilon_0}$ and the $k$ is normalized by $L= W_{y}$.
}}
\end{figure}

\begin{figure}[ht]
\begin{center}
\includegraphics{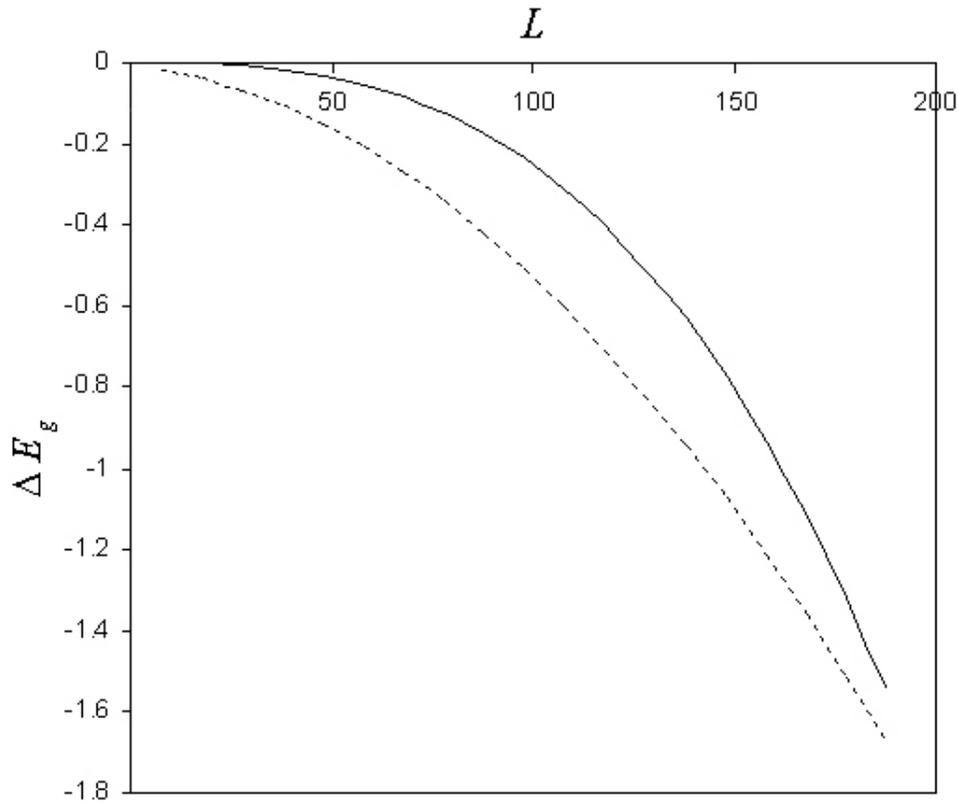}

\end{center}
\vspace{10 cm}
 \caption{\small {Calculated band gap renormalization in H-Shaped
  quantum wire within dynamical random phase(GW) approximation(upper curve)
  and quasi-static plasmon-pole approximation(lower curve).
  The screened potential is normalized by $\frac{2e^2}{\epsilon_0}$
  and the $k$ is normalized by $L= W_{y}$.}}
 \label{Fig:7}
\end{figure}
\end{document}